\documentclass[11pt]{article}

\usepackage[english]{babel}

\usepackage{amsmath}
\usepackage{amssymb}
\usepackage{amsthm}
\usepackage{amsfonts}
\usepackage{xcolor}
\usepackage{thmtools}  
\usepackage{hyperref}  

\declaretheorem[name=Theorem,numberwithin=section]{theorem}
\declaretheorem[name=Corollary,numberlike=theorem]{corollary}
\declaretheorem[name=Lemma,numberlike=theorem]{lemma}

\makeatletter
\providecommand*{\cupdot}{%
	\mathbin{%
		\mathpalette\@cupdot{}%
	}%
}
\newcommand*{\@cupdot}[2]{%
	\ooalign{%
		$\m@th#1\cup$\cr
		\hidewidth$\m@th#1\cdot$\hidewidth
	}%
}
\makeatother

\newcommand{\overbar}[1]{\mkern 1.5mu\overline{\mkern-1.5mu#1\mkern-1.5mu}\mkern 1.5mu}
\renewcommand{\Pr}{\operatorname{Pr}}
\newcommand{\E}{\mathbb{E}}

\theoremstyle{remark}
\newtheorem*{remark}{Remark}

\title{Concentration and maximin fair allocations for subadditive valuations}
\author{Uriel Feige\thanks{Weizmann Institute, Israel. {\tt uriel.feige@weizmann.ac.il}} \ and Shengyu Huang\thanks{EPFL, Switzerland. {\tt shhuang.huangsh@gmail.com}}}

\begin{document}
	
\maketitle

\begin{abstract}
  We consider fair allocation of $m$ indivisible items to $n$ agents of equal entitlements, with submodular valuation functions. Previously, Seddighin and Seddighin [{\em Artificial Intelligence} 2024] proved the existence of allocations that offer each agent at least a $\frac{1}{c \log n \log\log n}$ fraction of her maximin share (MMS), where $c$ is some large constant (over 1000, in their work). We modify their algorithm and improve its analysis, improving the ratio to $\frac{1}{14 \log n}$. 
  
  Some of our improvement stems from tighter analysis of concentration properties for the value of any subadditive valuation function $v$, when considering a set $S' \subseteq S$ of items, where each item of $S$ is included in $S'$ independently at random (with possibly different probabilities). In particular, we prove that up to less than the value of one item, the median value of $v(S')$, denoted by $M$, is at least two-thirds of the expected value, $M \geq \frac{2}{3}\E[v(S')] - \frac{11}{12}\max_{e \in S} v(e)$.
\end{abstract}

\section{Introduction}\label{sec:intro}

Dividing a set of resources among multiple agents is a common challenge in various real-world scenarios, such as divorce settlements, electronic frequency allocation, allocation of housing units, or scheduling medical appointments. The goal of fair division is to ensure that the distribution is perceived as fair by the agents involved. 




In this paper, we consider fair allocation of $m$ indivisible items to $n$ agents of equal entitlements, where each agent $i$ has a valuation function $v_i$. The fairness notion we adopt is maximin-share (MMS) fairness. Intuitively, MMS fairness can be seen as a generalization of the cut-and-choose protocol for multiple players. In a cut-and-choose protocol, one agent (``the cutter'') divides the items into two bundles, and the other agent (``the chooser'') selects one bundle she prefers. {One can} extend the idea {to $n$ agents, where $n$ is greater than two.} One agent acts as the cutter, and she gets to partition the items into $n$ bundles. The rest of the $n-1$ agents are then entitled to choose their bundles before the cutter. The cutter might get the bundle that she views as least valuable  for her. Thus a risk averse cutter is incentivized to cut in a way that maximizes the value of the least valuable bundle. Such a cut is referred to as an \textit{MMS partition}. 

Formally, the \textit{MMS value} of an agent {$a_i$} is defined as 
$$MMS_i^n(\mathcal{M}) = \max_{(B_1, ..., B_n) \in \Pi_n(\mathcal{M})} \min_{j \in [n]} v_i(B_j),$$
where 
$\mathcal{M}$ is the set of $m$ indivisible goods,  $\Pi_n(\mathcal{M})$ is the set of all possible $n$ partitions of the items, and $v_i$ corresponds to the valuation function of agent $a_i$. An allocation is considered an MMS allocation if it ensures that every agent receives a bundle worth at least her MMS value.

In this paper, we assume that $v_i$ is a \textit{monotone} and \textit{subadditive} function for all $1 \leq i \leq n$. 
\begin{itemize}
	\item \textbf{Monotonicity}. $v_i(S) \leq v_i(S') \;\;\;\; \forall S \subseteq S'$
	\item \textbf{Subadditivity}. $v_i(S \cup T) \leq v_i(S) + v_i (T) \;\;\;\; \forall S, T \subseteq \mathcal{M}$
\end{itemize}

	
Subadditive valuations are also known as complement-free valuations. In a complementary (superadditive) setting, items can have strong positive synergy and owning these item together can lead to higher value than the sum of their separate values. For example, a pair of shoes is typically far more valuable than a left or right shoe alone. Subadditive valuations exclude such scenarios while still encompassing many relevant real-world examples.
	
	Some well studied subclasses of the complement free hierarchy include: 
	\begin{itemize}
		\item Additive. $v(S) = \sum_{e \in S} v(e)$
		\item Submodular. $v(S \cap T) + v(S \cup T) \leq v(S) + v(T) \;\;\; \forall S, T$. Equivalently, $v$ is a submodular function iff for every item $j$ and $T \subset S$, $v(j \cup S) - v(S) \leq v(j \cup T) - v(T)$. The latter definition characterizes the diminishing returns property of submodular functions.
		\item Fractionally subadditive (XOS). $v(S) \leq \sum \alpha_i v(T_i)$ ($0 \leq \alpha_i \leq 1$) if for every item $j \in S$, $\sum_{i|j \in T_i} \alpha_i \geq 1$. Equivalently, $v$ is an XOS function if there is a finite collection of additive functions $v_1, v_2, ...$ such that for every set of items $S$, $v(S) = \max_j \{v_j(S)\}$. The equivalence of these two definitions is proved in~\cite{feige2009maximizing}.
	\end{itemize}
	
	It is worth noting that every additive function is submodular, every submodular function is fractionally subadditive, and every fractionally subadditive function is subadditive. 

For $\rho > 0$, a $\rho$-MMS allocation is one in which every agent gets a bundle of value at least a $\rho$-fraction of her MMS. Our goal is to find the highest value of $\rho$, that we refer to as $\rho_{SA}$, such that if every agent has a subadditive valuation, then there are $\rho_{SA}$-MMS allocations. Previously, Ghodsi et al.~\cite{ghodsi2022fair} has given an example establishing $\rho_{SA} \le \frac{1}{2}$, and also showed that $\rho_{SA} \ge \frac{1}{\lceil 10 \log m \rceil}$. Seddighin and Seddighin~\cite{seddighin2024improved} later showed that $\rho_{SA} \ge \Omega(\frac{1}{\log n \log \log n})$. Specifically, they gave a randomized allocation procedure in which with positive probability every agent gets a bundle worth at least $\frac{1}{1536 \log n \log (1536 \log n)}$ of her MMS value (see Theorem 4.2 in~\cite{seddighin2024improved}). 

By modifying the  allocation procedure of~\cite{seddighin2024improved} and improving its analysis, we obtain the following result.

\begin{restatable}{theorem}{thmmain}\label{thm:main}
	In every allocation instance in which agents have monotone subadditive valuations, there exists an allocation in which each agent obtains a bundle worth at least $\frac{1}{14 \log n}$ of her MMS value.
\end{restatable} 


At a high level, there are {three} main steps in our algorithm: {preprocessing,} tentative allocation, and uniform contention resolution. {The preprocessing step removes large items. Specifically, if an agent values an item more than $\frac{1}{14 \log n}$, we assign this item to this agent and remove them from the problem instance. We repeat this process until no agent values any item more than $\frac{1}{14 \log n}$.} Then, we apply the rounding technique by Feige~\cite{feige2009maximizing} to $\Theta(\log n)$ copies of the problem instance independently. This will give a tentative allocation with the caveat that an item might be allocated to multiple agents across all copies of the problem instance. To ensure a valid allocation, if an item $e \in \mathcal{M}$ is allocated to multiple agents, we select one of these agents with a probability proportional to the number of copies of $e$ that the agent holds and allocate $e$ to her. We argue that after this contention resolution procedure, each agent fails to hold a bundle worth $\frac{1}{14 \log n}$ of her MMS value with probability less than 
$\frac{1}{n}$. By the union bound, this implies we can find a $\frac{1}{14\log n}$-MMS allocation with positive probability, proving the existence of such an allocation.



In our analysis of the contention resolution procedure, we make use of earlier work of Talagrand (see Theorem~\ref{thm:talagrand}) to derive and make use of the following proposition. (Previous work, such as~\cite{seddighin2024improved}, made use of related but weaker propositions.) 

\begin{restatable}{proposition}{propupperbound}
	\label{prop:upper-bound}
	Let $v$ be a monotone and subadditive function. Given a set $S$, we construct $S'$ by independently sampling each element $e \in S$. If $v(e) \leq b$ for all $e \in S$, then $\E[v(S')] \leq \frac{3}{2}M + \frac{11}{8}b$, where $M$ is the median of $v(S')$.
\end{restatable} 

In Section~\ref{sec:tight}, we give for arbitrarily large $M$ a subadditive function $v$ for which $\E[v(S')] \simeq \frac{3}{2}M$ and $v(e) \leq 1$ for all $e \in S$. Since $M$ can be arbitrarily large, this example shows that the upper bound $\E[v(S')] \leq \frac{3}{2}M+ \frac{11}{8}b$ is nearly tight when $M$ is much larger than $b$. 

\subsection{Related work}


Fair division tackles the general problem of dividing limited resources among multiple agents. The fairness notion we look at in this paper is maximin-share (MMS) fairness, first introduced by Budish~\cite{budish2011combinatorial}. 

Extensive studies have been done for MMS allocation problems when agents have additive valuation functions. {Kurokawa, Procaccia, and Wang}~\cite{kurokawa2018fair}  gave an example showing that exact MMS allocations may not exist for general additive valuations when there are more than two agents. In the same paper~\cite{kurokawa2018fair}, they presented an algorithm to find a $\frac{2}{3}$-MMS allocation. {After a long sequence of works~\cite{amanatidis2017approximation,  ghodsi2022fair, garg2019approximating, garg2020improved, barman2020approximation, akrami2023simplification, akrami2024breaking}, the current best approximation for additive valuations is $\frac{3}{4}+\frac{3}{3836}$, achieved by Akrami and Garg~\cite{akrami2024breaking}.} 

For valuations beyond additive functions, Ghodsi et al.~\cite{ghodsi2022fair} gave counterexamples for obtaining an approximation ratio better than $\frac{3}{4}$ for submodular valuations and better than $\frac{1}{2}$ for fractionally subadditive valuations. They proved in the same paper the existence of a $\frac{1}{3}$-MMS allocation for submodular valuations and a $\frac{1}{5}$-MMS allocation for fractionally subadditive functions. Akrami et al.~\cite{akrami2024breaking} improved the approximation ratio for fractionally subadditive valuations to $\frac{3}{13}$.

More relevant to our work, Ghodsi et al.~\cite{ghodsi2022fair} showed the existence of $\frac{1}{10 \lceil \log m \rceil}$-MMS allocations for subadditive valuations. 
Seddighin and Seddighin~\cite{seddighin2024improved} later showed the existence of  $\frac{1}{1536 \log n \log (1536\log n)}$-MMS allocations.  We modify their approach and analysis and improve this approximation ratio to $\frac{1}{14 \log n}$.

Our analysis involves concentration bounds for subadditive functions. One such concentration bound, Lemma~\ref{lem:schechtman}, is a variation of Corollary 12 in~\cite{schechtman2003concentration}. Vondrák referred to that Corollary in his notes (see Theorem 4.1 in~\cite{vondrak2010note}), where he discussed the concentration properties of self-bounding functions (including non-negative submodular and fractionally subadditive functions with marginal values in $[0, 1]$). While Chernoff-strength concentration holds when a function $f: \{0, 1\}^n \rightarrow \mathbb{R}^+$ is submodular or fractionally subadditive, it fails to hold when $f$ is subadditive. Nevertheless, subadditive functions still have some concentration properties that do not hold for arbitrary $1$-Lipschitz functions. Proposition~\ref{prop:upper-bound}, Lemma~\ref{lem:concentration}, and Lemma~\ref{lem:schechtman} demonstrate some concentration properties for subadditive functions. 

\section{The allocation procedure}\label{sec:allocation-procedure}


Consider the following standard linear progamming (LP) relaxation for
combinatorial auctions, used by Dobzinski, Nisan, and Schapira~\cite{dobzinski2010approximation}. The indicator variable $x_{i, S}$ in the LP specifies whether agent $a_i$ gets set $S$.

\begin{equation*}
	\begin{array}{ll@{}ll}
		\text{maximize}  
		& \sum_{i, S} x_{i, S} \cdot v_i(S)&\\[5pt]
		\text{subject to}
		& \sum_{i, S|j \in S}   x_{i, S} \leq 1  & \;\;\;\textit{for every item $j$} \\[5pt]
		& \sum_{S}   x_{i, S} \leq 1  &\;\;\;\textit{for every agent $a_i$} \\[5pt]
		&    x_{i, S} \geq 0
	\end{array}
\end{equation*}

Given the MMS partitions of all agents, we set $x_{i, S_i^j} = \frac{1}{n}$ if $S_i^j$ belongs to $a_i$'s MMS partition and $x_{i, S} = 0$ otherwise. This encoding yields a feasible solution to the configuration LP above. By using the rounding technique by Feige for subadditive valuation functions~\cite{feige2009maximizing}, each agent will obtain a bundle worth half of her MMS value with probability at least $\frac{1}{2}$, as stated below.

\begin{lemma}[Rounding lemma]\label{lem:maximize-welfare}
	Given the MMS partitions of all agents, we can encode them into a feasible solution of the configuration LP described above. The rounding technique by Feige~\cite{feige2009maximizing} will give each agent a bundle worth at least half of her MMS value with probability at least $\frac{1}{2}$, assuming all agents have subadditive and monotone valuation functions.
\end{lemma}

We will make use of the rounding lemma above and the following concentration lemma to analyze our randomized algorithm. We defer the proof of the concentration lemma to Section~\ref{sec:concentration-lemma}. 

\begin{restatable}[Concentration lemma]{lemma}{lemconcentration}\label{lem:concentration}
	Let $v$ be a monotone and subadditive function. Given a set $S$, we construct $S'$ by independently sampling each element $e \in S$ with a probability of at least $\frac{1}{t}$, where $t \in \mathbb{N}^{+}$. 
	If $v(e) \leq \frac{8}{23t} \cdot v(S)$ for all $e \in S$, then $\Pr[v(S') \geq \frac{8}{23t} \cdot v(S)] \geq \frac{1}{2}$.
\end{restatable}

Our goal is to allocate $m$ indivisible items to $n$ agents so that every agent gets a bundle as close to their MMS value as possible. Each agent has her own valuation function that is assumed to be subadditive and monotone. Without loss of generality, assume that the MMS values of all agents are one. This can be done by scaling the valuation function of each agent. 

Our randomized algorithm below allocates a bundle worth at least $\frac{1}{14 \log n}$ to each agent with probability at least $1 - \frac{1}{n}$. In this algorithm, we will make $t$ copies of the problem instance, where $t$ is a fixed parameter that will be specified later.


\begin{enumerate}
	\item \textbf{Preprocessing.} {If an item is worth more than $\frac{4}{23t}$ to some agent, we allocate this item to this agent.} Then we remove this agent and this item. Repeat this process until no item is worth more than $\frac{4}{23t}$ {for any agent}. 
	\item \textbf{Tentative allocation.} Apply the rounding lemma (Lemma~\ref{lem:maximize-welfare}) independently to $t=\frac{56}{23} \log n$ copies of the problem instance $I(\mathcal{N}, \mathcal{M}, \mathcal{V})$. Every agent holds one bundle in each copy, so every agent has $t$ bundles across all $t$ copies. The \textit{$t$-tentative} bundle of an agent is the union of these $t$ bundles. Since {an} item may appear in the $t$-tentative bundles of different agents, the allocation formed by $t$-tentative bundles is not yet valid.
	\item \textbf{Uniform contention resolution.} For every item $e \in \mathcal{M}$, if it is allocated to more than one agent, we uniformly at random pick one copy of $e$ and assign $e$ to the agent that holds this copy. After this step, each item is assigned to at most one agent. 
	
\end{enumerate}

Note that removing $k$ items and $k$ agents does not decrease the MMS value of the remaining agents (see for example Proposition 3 and Corollary 4 in~\cite{garg2019approximating}). During the preprocessing step, if the MMS value of some remaining agent becomes strictly greater than one, we scale her valuation function so that her MMS value always stays to be one. 

The following lemma implies that for $t = \frac{56}{23} \log n$~\footnote{If $\frac{56}{23} \log n$ is not an integer, we make $\lceil \frac{56}{23} \log n \rceil$ copies instead. We omit the ceiling function for ease of reading.}, an agent fails to obtain a bundle worth more than $\frac{4}{23} \cdot \frac{23}{56 \log n} = \frac{1}{14 \log n}$ with probability at most $(\frac{3}{4})^{\frac{56}{23} \log n} < \frac{1}{n}$.  By the union bound, all agents can obtain a bundle worth at least $\frac{1}{14 \log n}$ with positive probability. The agents removed during the preprocessing step have obtained a bundle (that contains only one single item) worth at least $\frac{1}{14 \log n}$. Hence, our allocation procedure gives a $\frac{1}{14 \log n}$-MMS allocation.

\begin{lemma}\label{lem:fixed-agent}
		For any fixed agent $a_i \in \mathcal{N}$, the probability of her obtaining a bundle worth less than $\frac{4}{23t}$ is at most $(\frac{3}{4})^t$.
\end{lemma}
\begin{proof}		
	We first fix an arbitrary ordering of the $t$ problem instances for all agents. Then we inductively argue that the union of the first $k$ bundles of $a_i$ is worth less than $\frac{4}{23t}$ with probability at most $(\frac{3}{4})^k$. This implies that the final bundle that $a_i$ obtains, which is the union of her $t$ bundles after uniform contention resolution, is worth less than $\frac{4}{23t}$ with probability at most $(\frac{3}{4})^t$.
	
	Formally speaking, let $B_i(I_j)$ denote the bundle agent $a_i$ obtains in the $j$-th problem instance and let $\tilde{B}_i(I_j)$ denote the remaining subset of $B_i(I_j)$ after uniform contention resolution. For  $1 \leq k \leq t$, we use $E_k$ to denote the event that $v_i(\cup_{j=1}^{k} \tilde{B}_i(I_j)) < \frac{4}{23t}$. We will prove by induction that for every $1 \leq k \leq t$, $\Pr[E_{k}] \leq (\frac{3}{4})^k$. Therefore, $\Pr[E_t] = \Pr[\cup_{j=1}^{t} \tilde{B}_i(I_j) < \frac{4}{23t}] \leq (\frac{3}{4})^t$. Since $\cup_{j=1}^{t} \tilde{B}_i(I_j)$ is the bundle $a_i$ obtains in the end, this implies the probability of $a_i$ obtaining a bundle worth more than $\frac{4}{23t}$ is at least $1 - \frac{1}{n}$.
	
	For any $1 \leq k < t$, $\Pr[E_{k+1} | \overbar{E}_{k}] = 0$. By the law of total probability,
	\begin{align*}
		\Pr[E_{k+1}] = \Pr[E_{k+1} | E_{k}] \Pr[E_{k}].
	\end{align*}
	To complete our inductive proof and show $\Pr[E_t] \leq (\frac{3}{4})^t$, it suffices to show $\Pr[E_1] \leq \frac{3}{4}$ and $\Pr[E_{k+1} | E_k] \leq \frac{3}{4}$. 
	
	\textbf{Base case.} We show $\Pr[E_1] \leq \frac{3}{4}$ by showing $\Pr[\overbar{E}_1] \geq \frac{1}{4}$. Assuming $v_i(B_i(I_1))$ is greater than $\frac{1}{2}$, we can apply the concentration lemma (Lemma~\ref{lem:concentration}) since 
	\begin{itemize}
		\item every element is smaller than $\frac{4}{23t} \leq \frac{8}{23t} \cdot v_i(B_i(I_1))$ after the preprocessing step;
		\item every element $e \in B_i(I_1)$ is sampled independently with probability $\frac{1}{m(e)}$, where $m(e) \leq t$~\footnote{Note that the rounding technique from~\cite{feige2009maximizing} might not allocate all items in one problem instance, so $m(e)$ could be strictly smaller than $t$.} is the number of occurrences of $e$ across $t$ copies of problem instances. 
	\end{itemize}
	
	Hence, assuming $v_i(B_i(I_1)) \geq \frac{1}{2}$, $v_i(\tilde{B}_i(I_1)) \geq \frac{4}{23t}$ with probability at least $\frac{1}{2}$. Together with the rounding lemma, it implies $\Pr[\overbar{E}_1] \geq \frac{1}{4}$.
	
	\textbf{Inductive step.} We now show
	$\Pr[E_{k+1}] \leq (\frac{3}{4})^{k+1}$, assuming $\Pr[E_k] \leq (\frac{3}{4})^k$ for some $1 \leq k < t$. For every $e$ that is in the $(k+1)$-th bundle of $a_i$, if $e$ already exists in one of $a_i$'s first $k$ bundles and has been assigned to $a_i$ during uniform contention resolution, then $e$ will be in the union of $a_i$'s first $k+1$ bundles. Otherwise, $e$ would be included into the union of $a_i$'s first $k+1$ bundles with probability $\frac{1}{m(e) - m_i^k(e)}$, where $m_i^k(e)$ is the number of occurrences of $e$ across $a_i$'s first $k$ bundles. 
	
	Formally, for every $e \in B_i(I_{k+1})$, if $e \in \cup_{j=1}^k \tilde{B}_i(I_j)$, $\Pr[e \in \cup_{j=1}^{k+1} \tilde{B}_i(I_j)] = 1$. If $e \not\in\cup_{j=1}^k \tilde{B}_i(I_j)$, $\Pr[e \in \cup_{j=1}^{k+1} \tilde{B}_i(I_j)] = \frac{1}{m(e) - m_i^k(e)}$. Hence, we can still apply the rounding lemma and the concentration lemma in a similar way and conclude $\Pr[\overbar{E}_{k+1} | E_{k}] \geq \frac{1}{4}$.
	
\end{proof}

\section{Concentration for subadditive valuations}\label{sec:concentration-lemma}

In this section, we will prove the following concentration lemma that we used in Section~\ref{sec:allocation-procedure}.

\lemconcentration*

Let $M$ be the median of $v(S')$. That is, $\Pr[v(S') \leq M] \geq \frac{1}{2}$ and $\Pr[v(S') \geq M] \geq \frac{1}{2}$. The high-level idea is to show $\frac{v(S)}{t} \leq \E[v(S')] \leq \frac{23}{8} M$, assuming no single item is worth more than $M$. This implies $M \geq \frac{8}{23} \cdot \frac{v(S)}{t}$ and $\Pr[v(S') < \frac{8}{23} \cdot \frac{v(S)}{t}] \leq \Pr[v(S') < M] \le \frac{1}{2}$.

\subsection{A lower bound on $\E[v(S')]$}\label{sec:lower-bound}


When $t$ is an integer, the bound $\E[v(S')] \geq \frac{v(S)}{t}$ is known (see for example Proposition 2.2 in~\cite{feige2009maximizing}), but we briefly reproduce the proof here for completeness.

We partition the set $S$ into $t$ subsets by placing every element $e \in S$ into one of these $t$ subsets uniformly at random. Our sampled subset $S'$ can be obtained by selecting one of these $t$ subsets uniformly at random, so $\E[v(S')] = \frac{1}{t} \sum_{i=1}^t v(S_i) \geq \frac{v(S)}{t}$, where the last step holds because subadditivity implies $\sum_{i=1}^t v(S_i) \geq v(S)$.

Note that this lower bound might fail to hold if $t$ is not an integer. That is, $\E[v(S')]$ could be strictly smaller than $p \cdot v(S)$ if we sample each element with probability $p$, where $0 < p < 1$. Consider an example when $v(\emptyset) = 0$, $v(S') = 1$ for all $S' \subsetneq S$,  and $v(S) = 2$. For $s = |S|$, $\E[v(S')] = 2 p^s + (1 - (1-p)^s - p^s) = 1 - (1-p)^s + p^s$. However, $1 - (1-p)^s + p^s \geq 2p$ does not hold for $\frac{1}{2} < p < 1$ and $s$ sufficiently large. 

\subsection{An upper bound on $\E[v(S')]$}\label{sec:upper-bound}
Let $M$ be the median of $v(S')$. We will show $\E[v(S')] \leq \frac{23}{8} M$, assuming no single item is worth more than $M$. 

The high-level idea is to partition the range of $v(S')$ into infinitely many disjoint intervals $I_0, I_1, ...$, and then maxmize $\sum_{i \geq 0} \Pr[I_i] \sup(I_i)$. Assume $b$ is the maximum value of a single item. For some $q \in \mathbb{N}^{\geq 2}$, we partition the range into $(0, M]$, $(M, qM]$, $(qM, qM+b]$, $(qM+b, qM+2b]$, etc. Apart from the first two intervals $(0, M]$ and $(M, qM]$, all the other intervals can be represented by $(qM+kb, qM+(k+1)b]$ for some $k \in \mathbb{N}$. We will then upper bound $\E[v(S')]$ by showing that the upper bound on $\Pr[v(S') > qM+kb]$ decreases exponentially with $k$. 

One way to upper bound $\Pr[v(S') > qM+kb]$ is to use the following lemma. Lemma~\ref{lem:schechtman} follows directly from Lemma~\ref{lem:schechtman-intermediate} and Corollary~\ref{cor:talagrand} that will be proved later.

\begin{restatable}{lemma}{lemschechtman}
	\label{lem:schechtman}
	Let $f: \{0, 1\}^n \rightarrow {\mathbb{R}}$ be a monotone and subadditive function that satisfies $f({\textbf{0}}) = 0$ and $|f(x) - f(y)| \leq b \cdot h(x, y)$ for all $x, y \in \{0, 1\}^n$, where $h(x, y) = |\{i \;|\; x_i \neq y_i\}|$. 
	Then, for all $c_i > 0$,  $k  \in \mathbb{N}$, and $q \in \mathbb{N}^+$,
	\begin{align*}
		\Pr[f(x) > \sum_{i=1}^q c_i +kb] \leq \frac{q^{-k-1}}{\Pi_{i=1}^q \Pr[f(x) \leq c_i]}.
	\end{align*}
\end{restatable}

Lemma~\ref{lem:schechtman} offers somewhat stronger bounds than Corollary 12 in~\cite{schechtman2003concentration}, but restricted to domain $\{0, 1\}^n$. Seddighin and Seddighin used Corollary 12 in~\cite{schechtman2003concentration} to derive an upper bound of $\E[v(S')]$ (see the proof of Lemma 4.1 in~\cite{seddighin2024improved}). 

\begin{remark}
	Corollary 12 in~\cite{schechtman2003concentration} is stated by letting $c_1 = ... = c_q$. We note here that the more general form in Lemma~\ref{lem:schechtman} may give a better upper bound. Suppose for some montotone and subadditive function $f$, we have $\Pr[f(x) \leq M] = \frac{1}{2}$ and $\Pr[f(x) \leq 2M] = \frac{2}{3}$. One can apply Lemma~\ref{lem:schechtman} to obtain the upper bound of $\Pr[f(x) > 3M + b]$ in two different ways.
	\begin{enumerate}
		\item Choose $q = 2$ and let $c_1 = M$ and $c_2 = 2M$. $$\Pr[f(x) > 3M + b] \leq 2^{-2} \cdot \Pr[f(x) \leq M]^{-1} \cdot \Pr[f(x) \leq 2M]^{-1} = \frac{3}{4}.$$
		\item Choose $q = 3$ and let $c_1 = c_2 = c_3 = M$. $$\Pr[f(x) > 3M + b] \leq 3^{-2} \cdot \Pr[f(x) \leq M]^{-3} = \frac{8}{9}.$$
	\end{enumerate}
\end{remark}

Instead of using Corollary 12 in~\cite{schechtman2003concentration} (or Lemma~\ref{lem:schechtman}) to upper bound $\E[v(S')]$, we adopted a slightly different approach by observing that Corollary 12 in~\cite{schechtman2003concentration} (or Lemma~\ref{lem:schechtman}) is a direct consequence of Theorem 3.1.1 in~\cite{talagrand1995concentration}, which is reproduced below.

\begin{theorem}\label{thm:talagrand}
	Consider a product probability space $(\Omega = \Pi_{i=1}^n \Omega_i, P=\Pi_{i=1}^n \mu_i)$. Given $q \in \mathbb{N}^+$ and $x, y^1, ..., y^q \in \Omega$, we define the ``Hamming distance'' between $x$ and $(y^1, ..., y^q)$ by
	\begin{align*}
		h(x; y^1, ..., y^q) = |\{i; x_i \not\in\{y^1_i, ..., y^q_i\}\}|,
	\end{align*}
	and the distance between $x$ and $A_1, ..., A_q \subseteq \Omega$ is defined by
	\begin{align*}
		h(x; A_1, ..., A_q) = \inf\{h(x; y^1, ..., y^q); y^1 \in A_1, ..., y^q \in A_q \}.
	\end{align*}
	We have
	\begin{align*}
		\int q^{h(x; A_1, ..., A_q)} d\Pr(x) \leq \frac{1}{\Pi_{i=1}^q \Pr[A_i]}.
	\end{align*}
	In particular, we have
	\begin{align*}
		\Pr[{h(x; A, ..., A) \geq k}]\leq \Pr[A]^{-q}q^{-k}.
	\end{align*}
\end{theorem}

In the context of our problem, valuation functions are set functions, so we can restrict $\Omega$ to be $\{0, 1\}^n$. We interpret bitstrings and their representing sets interchangeably.  The Hamming distance defined above between $x \in \{0, 1\}^n$ and $A_1, ..., A_q \subseteq \{0, 1\}^n$ would be
\begin{align*}
	h(x; A_1, ..., A_q) 
	= & \min \{h(x; y^1, ..., y^q); y^1 \in A_1, ..., y^q \in A_q\} \\ 
	= & |x \setminus \bigcup_{1 \leq i \leq q} y_i | + | \bigcap_{1 \leq i \leq q} y_i \setminus x |.
\end{align*}


Since our application only concerns $\Omega = \{0, 1\}^n$, we reformulate Theorem~\ref{thm:talagrand} and use the following corollary to prove Lemma~\ref{lem:schechtman} and Lemma~\ref{lem:schechtman-intermediate}.

\begin{corollary}
	\label{cor:talagrand}
	For $x \in \{0, 1\}^n$, $A_i \subseteq \{0, 1\}^n$ and every $q \in \mathbb{N}^+$, we have
	
	$$\sum_{x\in \{0, 1\}^n} q^{h(x; A_1, ..., A_q)} \Pr[x] \leq \frac{1}{\Pi_{i=1}^q \Pr[A_i]}.$$
	
	In particular, we have for every $k \in \mathbb{N}$,
	
	$$\Pr[h(x; A_1, ..., A_q) > k] \leq \frac{q^{-k-1}}{\Pi_{i=1}^q \Pr[A_i]}.$$
\end{corollary}

\begin{proof}
	The integral in Theorem~\ref{thm:talagrand} can be replaced by a sum, because the domain is discrete and finite. 
	
	The proof for the ``in particular'' part in   Corollary~\ref{cor:talagrand} can be derived from the ``in particular" part in Theorem~\ref{thm:talagrand}. For completeness, we show how the ``in particular" part of the corollary follows from its main part. The condition $h(x; A_1, ..., A_q) > k$ implies that $\frac{q^{h(x; A_1, ..., A_q)}}{q^{k+1}} \geq 1$ for all $q$, since $h(x; A_1, ..., A_q)$ can only take integer values, 
	$h(x; A_1, ..., A_q) > k$ is equivalent to $h(x; A_1, ..., A_q) \ge k+1$. Hence, 
	\begin{align*}
		\Pr[h(x; A_1, ..., A_q) > k] 
		= & \sum_{x \in \{0, 1\}^n} {\textbf{1}}_{h(x; A_1, ..., A_q) > k} \Pr[x] 
		\leq  \sum_{x \in \{0,1\}^n} \frac{q^{h(x; A_1, ..., A_q)}}{q^{k+1}} \Pr[x] \\
		\leq & q^{-k-1} \cdot \sum_{x\in \{0, 1\}^n} q^{h(x; A_1, ..., A_q)} \Pr[x] \\ 
		\leq & \frac{q^{-k-1}}{\Pi_{i=1}^q \Pr[A_i]},
	\end{align*}
	where the last inequality uses the main part of Corollary~\ref{cor:talagrand}.
\end{proof}

The following lemma and Corollary~\ref{cor:talagrand} together imply Lemma~\ref{lem:schechtman}.

\begin{lemma}\label{lem:schechtman-intermediate}
	Let $A_i = \{y^i \in \{0, 1\}^n \;|\; f(y^i) \leq c_i\}$. Then for any $k \in \mathbb{N}$,
	\begin{align*}
		\{x \;|\; f(x) > \sum_{i=1}^q c_i + kb\} \subseteq \{x \;|\; h(x; A_1, ..., A_q) > k\},
	\end{align*}
\end{lemma}
\begin{proof}
	Assume for contradiction $h(x; y^1, ..., y^q) \leq k$ for some $x, y^1, ..., y^q \in \{0, 1\}^n$, where $f(y^i) \leq c_i$ for every $1 \leq i \leq q$. We show that it implies $f(x) \leq \sum_{i=1}^q c_i + kb$.
	
	Let $p_{\bar{x}}$ be the subset of $x$ that is not covered by all $y^i$, i.e., $p_{\bar{x}} = x \setminus \bigcup_{i=1}^q y^i$. Let $p_{x, i}$ be the intersection between $x$ and $y^i$, i.e.,  $p_{x, i} = x \cap y^i$ for $1 \leq i \leq q$. By subadditivity and $x = p_{\bar{x}} \cup \bigcup_{i=1}^q p_{x, i}$, we have $f(x) \leq f(p_{\bar{x}}) + \sum_{i=1}^q f(p_{x, i})$.
	
	$f(p_{x, i}) \leq c_i$ for all $1 \leq i \leq q$ since $f(p_{x, i}) \leq f(y^i) \leq c_i$, where the first inequality comes from monotonicity, and the second inequality comes from our assumption that $f(y^i) \leq c_i$ for all $1 \leq i \leq q$. It follows that $\sum_{i=1}^q f(p_{x, i}) \leq \sum_{i=1}^q c_i$.
	
	$f(p_{\bar{x}}) \leq kb$ because $|f(x) - f(y)| \leq b \cdot h(x, y)$ for all $x, y \in \{0, 1\}^n$, and $|f(p_{\bar{x}}) - f({\textbf{0}})| \leq b \cdot h(p_{\bar{x}}, {\textbf{0}})$ in particular. It follows that $h(p_{\bar{x}}, {\textbf{0}}) = |p_{\bar{x}}| \leq h(x; y^1, ..., y^q) \leq k$.
\end{proof}


Now we will use the main part of Corollary~\ref{cor:talagrand} and Lemma~\ref{lem:schechtman-intermediate} to upper bound $\E[v(S')]$. Recall that $S'$ is our sampled subset of $S$ and our valuation function $v$ is monotone and subadditive. Let $A = \{y \in \{0, 1\}^{|S|} \;|\; v(y) \leq M\}$. Lemma~\ref{lem:schechtman-intermediate} implies that if $h(S'; A, ..., A) = k$, then $v(S') \leq qM + kb$. 
Let $h_k := \Pr[h(S'; A, ..., A) = k]$ and $H(S') := h(S'; A, ..., A)$.
\begin{align*}
	\E[v(S')]\leq & \Pr[A] M + (h_0 - \Pr[A]) qM + \sum_{k \geq 1} h_k  (qM + kb) \\
	= & \Pr[A] (1-q) M  + \sum_{k \geq 0} h_k (qM+kb) \\
	= & \Pr[A] (1-q) M  + qM \sum_{k \geq 0} h_k + b \sum_{k \geq 0} k h_k \\
	= & \Pr[A] (1-q) M + qM + \E[H(S')] b \\
	= & \; (\Pr[A] + (1- \Pr[A])q) \cdot M + \E[H(S')] \cdot b.
\end{align*}
For $q=2$, assuming $b=M$ and $\Pr[A] = \frac{1}{2}$ as before, it implies $\E[v(S')] \leq \frac{3}{2}M+\E[H(S')]b$.

To get an upper bound on $\E[H(S')] = \sum_{i\geq1} i \cdot h_i$, we can maximize it under two constraints: 1) $\sum_{i \geq 1} h_i \leq \frac{1}{2}$ and 2) $\sum_{i \geq 0} 2^i \cdot h_i \leq 4$. The first constraint comes from our assumption that $h_0 \geq \Pr[A] = \frac{1}{2}$, and the second constraint is imposed by the first part of Corollary~\ref{cor:talagrand}. 

The second constraint is equivalent to $\sum_{i \geq 1} 2^i \cdot h_i \leq \frac{7}{2}$, because $h_0 \geq \frac{1}{2}$. Suppose that there was some integer $t > 0$ such that $2^t \cdot \frac{1}{2} = \frac{7}{2}$. Then, convexity of the exponential function would imply that $\max \sum_{i \geq 1} i \cdot h_i = \frac{t}{2}$. However, since $t$ can only be an integer and $2^3 \cdot \frac{1}{2} > \frac{7}{2}$, we have to make both $h_3$ and $h_2$ non-zero such that $h_3 + h_2 = \frac{1}{2}$ and $2^2 \cdot (\frac{1}{2} - h_3) + 2^3 \cdot h_3 = \frac{7}{2}$. Solving this equation gives $h_3 = \frac{3}{8}$. This implies $\max \sum_{i \geq 1} i \cdot h_i = 2 \cdot (\frac{1}{2} - \frac{3}{8}) + 3 \cdot \frac{3}{8} = \frac{11}{8}$.

Putting all together, we have $\E[v(S')] \leq  \frac{3}{2}M + \frac{11}{8}b$, where $M$ is the median of $v(S')$ and $b$ is the value of the largest item. If $b=M$, $\E[v(S')] \leq \frac{23}{8}M$. That is, the expected value of the sampled subset will not be bigger than $\frac{23}{8}$ times its median.

The proof above shows Proposition~\ref{prop:upper-bound}. In addition, we have now developed all the tools needed to prove Theorem~\ref{thm:main}. 
\thmmain*
\begin{proof}
We choose $t = \frac{56}{23} \log n$~\footnote{If $\frac{56}{23} \log n$ is not an integer, we choose $t = \lceil \frac{56}{23} \log n \rceil$. We omit the ceiling function for ease of reading.} The agents removed during the preprocessing step have obtained a bundle (that contains only one single item) worth at least $\frac{4}{23} \cdot \frac{23}{56 \log n} =  \frac{1}{14 \log n}$. For the remaining agents, their MMS value did not decrease.
Lemma~\ref{lem:fixed-agent} then shows that an agent fails to obtain a bundle worth more than $\frac{4}{23} \cdot \frac{23}{56 \log n} = \frac{1}{14 \log n}$ with probability at most $(\frac{3}{4})^{\frac{56}{23} \log n} < \frac{1}{n}$.  By the union bound, with positive probability all agents obtain a bundle worth at least $\frac{1}{14 \log n}$.  

The proof of Lemma~\ref{lem:fixed-agent} made use of the rounding lemma (Lemma~\ref{lem:maximize-welfare}) and the concentration lemma (Lemma~\ref{lem:concentration}), were this latter lemma was proved in this section.
\end{proof}

\subsection{Tightness of the upper bound}
\label{sec:tight}

We show that the upper bound $\E[v(S')] \leq \frac{3}{2}M+ \frac{11}{8}b$ as stated in Proposition~\ref{prop:upper-bound} is nearly tight by giving a subadditive function $v$ that shows $\E[v(S')] \simeq \frac{3}{2}M$, where the median $M$ is an arbitrarily large integer, and where $b$, the value of the largest item, is $1$. Hence, $b$ is negligible compared to the median $M$.

Fix a large integer $M$ and consider a ground set with $s > 2M$ elements, with $s$ being an even integer. Sample each element with probability $\frac{1}{2}$. The subadditive function $v$ is defined as follows.

\[ v(S') = \begin{cases} 
	|S'| & |S'| \leq M \\
	M     & M < |S'| \leq \frac{s}{2} \\
	M+|S'|-\frac{s}{2} & \frac{s}{2} < |S'| \leq \frac{s}{2}+M \\
	2M   & \frac{s}{2}+M < |S'|
\end{cases}
\]

Due to the symmetry of the binomial coefficients, the median of $v(S')$ is $M$. 

We now calculate $\E[v(S')]$ by looking at the case when $0 \leq |S'| \leq s/2$ and when $s/2 + 1 \leq |S'| \leq n$.

\begin{align*}
	\E[v(S')] 
	= & 2^{-s} \cdot [\sum_{i=0}^{s/2} M{s \choose i} - \sum_{i=0}^M (M-i) {s \choose i}] + \\
	& 2^{-s} \cdot [\sum_{i=s/2+1}^s 2M{s \choose i} - \sum_{i=0}^M (M-i){s \choose \frac{s}{2}+i}] \\
	= & 2^{-s} \cdot [\sum_{i=0}^{s}M {s \choose i}+\sum_{i=s/2+1}^s 2M{s \choose i}] - \\
	& 2^{-s} \cdot [\sum_{i=0}^M (M-i) {s \choose i} + \sum_{i=0}^M (M-i){s \choose \frac{s}{2}+i}] \\
	= & \frac{3}{2}M - O(\frac{M^2}{\sqrt{s}})
\end{align*}

The last step is derived by lower bounding the term $M \cdot 2^{-s} \sum_{i=0}^M [{s \choose i} + {s \choose \frac{s}{2}+i}]$, which is greater than $2^{-s} [\sum_{i=0}^M (M-i){s \choose i} + \sum_{i=0}^M (M-i) {s \choose \frac{s}{2}+i}]$. For large $s$, the binomial distribution is well approximated by the normal distribution. That is,
\begin{align*}
	{s \choose i} 2^{-s} \approx \frac{1}{\sqrt{\frac{\pi s}{2}}} e^{-2(x-\frac{s}{2})^2/s}.
\end{align*}
Therefore, $2^{-s}\sum_{i=0}^M [{s \choose i} + {s \choose \frac{s}{2}+i}] = O(\frac{M}{\sqrt{s}})$ and $M \cdot 2^{-s}\sum_{i=0}^M [{s \choose i} + {s \choose \frac{s}{2}+i}] = O(\frac{M^2}{\sqrt{s}})$. 

By choosing $s$ sufficiently large compared to $M$, $\E[v(S')]$ can be made as close to $\frac{3}{2}M$ as desired.

\section{Discussion}

We have shown that $\frac{1}{14 \log n}$-MMS allocations exist for subadditive valuations. We have no reasonable doubt that this ratio can be improved.

Even without any change to the allocation algorithm, it may well be that its approximation ratio is better than $\frac{1}{14 \log n}$. Our analysis uses the bound $\E[v(S')] \leq \frac{3}{2}M + \frac{11}{8}b$ from Proposition~\ref{prop:upper-bound}. This bound is nearly tight when $b$ is much smaller than $M$, but we use it when $b = M$, in which case it is not known to be tight. Consequently, the analysis of our allocation algorithm is not known to be tight, and the leading constant in the approximation ratio may well be better than  $\frac{1}{14}$.
(However, even with an improved Proposition~\ref{prop:upper-bound}, the approximation ratio will remain worse than $\frac{1}{8 \log n}$. This is because there are examples in which $M = b = 1$ and $\E[v(S')] = 1.678$. This happens when $v$ is additive over $S$, $v(e) = 1$ for every $e \in S$, and each item is sampled with probability $\frac{1.678}{|S|}$.)

It is likely that much greater improvements in the approximation ratio can be obtained by changing the allocation algorithm. The analysis of the algorithm establishes that with positive probability, after contention resolution, for every agent at least one of her $t$ bundles has value at least $\frac{1}{14 \log n}$. If we would give each agent only the items that she received from this bundle, and not those not in this bundle, the (lower bound on the) approximation ratio will not change, but the vast majority of items will not be allocated at all. This suggests that the allocation algorithm is ``wasteful" in its use of items, and that better allocation algorithms should exist.


\subsection*{Acknowledgments}
		
		We thank Ola Svensson for useful comments on this work, and Gideon Schechtman for clarifying to us some of the previous work on concentration inequalities for subadditive functions. This research was supported in part by the Israel Science Foundation (grant No. 1122/22).

		\bibliographystyle{plain} 

	\end{document}